\documentstyle[12pt]{article}

\newlength{\signlength}

\renewcommand{\ge}{\settowidth{\signlength}{$>$}\setlength{\unitlength}{0.1\signlength}\mbox{\begin{picture}(17,7.5)\linethickness{0.045\signlength}\put(0.0,1){\makebox(17,7.5){$>$}}\multiput(4.5,-1)(0.25,0.125){32}{\line(1,0){0.25}}\end{picture}}}

\input{tcilatex}

\begin{document}

\title{2D XY Ordering in Striped $Cu O_2$ Planes in Underdoped $Bi_2 Sr_2 Ca Cu_2
O_{8+\delta}$ }
\author{G.G.Sergeeva}
\date{}
\maketitle

\begin{abstract}
It was discussed the possibility of determination of the temperature of 2D
XY ordering, $T_{BKT}$, in striped $Cu O_2$ planes for quasi-two dimensional
HTS with the semiconducting behavior of the out-of-plane resistivity $%
\rho_c(T)$ in normal state.  From the results of measurements $\rho_c(T)$
[1] for samples of underdoped $Bi_2 Sr_2 Ca Cu_2 O_{8+\delta}$ the values of 
$T_ {BKT}$ were found out [2]. Its are compared with results of measurements
of frequency dependent conductivity $\sigma(\omega) $[3]. It is shown that
these both results give close value of $T_ {BKT}\sim 62 K$ for samples with $%
T_c=74 K$, and verify the importance of taken into account two-dimensional
fluctuations effects at $T>T_c$.
\end{abstract}

\thispagestyle{plain}

\begin{center}
National science center "Kharkov Institute of Physics and Technology",
Academicheskaya st. 1, 61108, Kharkov, Ukraine
\end{center}

The hypothesis that the the semiconducting behavior of the out-of-plane
resistivity $\rho_c(T)$ in quasi-two dimensional HTS is the result of strong
two-dimensional antiferromagnetic (AFM) and superconducting fluctuations in
striped $Cu O_2$ planes (with metallic and insulating stripes) at $%
T_c<T<T^{*}$ was discussed in Refs. [2,4] (here $T^{*}$ is the charge
ordering temperature). In the region, where $\rho_c$ exceeds the Mott limit $%
\rho_M\sim 10^{-2} \Omega \cdot cm$, the charge transport along axis $%
\widehat{c}$ can be treated as a process of tunneling through a
nonconducting barrier with temperature depending probability $t_c(T)\sim
|\xi_c(T)/\xi_{ab}(T)|^{2}$: 
\begin{equation}  \label{1}
\rho_c \simeq \rho_M (N_0 t_{c})^{-1}
\end{equation}
where $N_0$ is the density states in the $CuO_2$ plane, $\xi_{ab}(T)$ and $%
\xi_c(T)$ are the correlations lengths in the $CuO_2$ plane and along axis $%
\widehat{c}$, respectively. The coexisting superconducting, dielectric and
metallic regions in striped $Cu O_2$ planes at $T_{c}<T< T^{*}$ leads to
different paths of out-of-plane charge tunneling and to different
temperature dependencies of the correlations lengths $\xi_{ab}(T)$ and $%
\xi_c(T)$. In the temperature regions $T \gg T_{BKT}$, where $\xi_c$ does
not depends on temparature and $\xi_{ab}(T)$ became formed by
Ginzburg-Landau dependence $\xi_{ab}(T)= \xi_{ab} (T/T_{BKT}-1)^{-1/2}$, it
follows [2]: 
\begin{equation}  \label{2}
\rho_c(T)= T_{BKT}\frac{\rho_M \xi_{ab}^{2}}{N_0 \xi_{c}^{2}}
(T-T_{BKT})^{-1},
\end{equation}
where $\xi_{ab} $ and $\xi_c$ are the values of the correlations lengths at $%
T_{BKT}$. Measurements of $\rho_c(T)$ [1] allows us to determine $T_{BKT}$
without any fitting parameter: 
\begin{equation}  \label{3}
T_{BKT}=\frac{\rho_i T_i - \rho_k T_k}{\rho_i - \rho_k},
\end{equation}
where pair points $T_i, T_k$ on curve $\rho_c(T)$ are taken approximately on
15 - 20K above the temperature, where semiconducting trend of $\rho_c$ end
off at low temperatures. The calculations of averaged values $\overline{
T_{BKT}}$ for four pairs points $T_i, T_k$ lead to the following results: $%
61.72 K,\hspace{5mm} 60.62K,\hspace{5mm} 56.2 K,\hspace{5mm}47.92 K $,
respectively for samples with $\delta$, which are equals $0.2135,\hspace{5mm}%
0.216,\hspace{5mm}0.22,\hspace{5mm}0.23$.

The dependence (2) with calculated $\overline{ T_{BKT}}$ well describes the
results of the measurements of out-of-plane resistivity for single crystalls
Bi-2212 [1] in normal state in enough big temperature interval, $96 K+\Delta
T_{fl} $, which is beginning at nearly undependent on oxygen concentration $%
\delta$ temperature $96 K$. $\Delta T_{fl} $ is depending on $\delta$ and is
changing from 24 K at $\delta=0.23$ up to 104 K at $\delta=0.2135$ [2]. It
is seen that values $T_{BKT}$ and $\Delta T_{fl}$ for samples are increasing
at decreasing oxygen concentration.

It is very interesting to compare these results with the measurements of
frequency dependent conductivity, $\sigma (\omega ,T)$, for epitaxial films
of underdoped $Bi_{2}Sr_{2}CaCu_{2}O_{8+\delta }$ with $T_{c}=74K$ (see
fig.2 from ref.[3]). The temperature of two-dimensional melting, $T_{BKT}$,
determines as the temperature of the crossover of two curves, namely $(\hbar
\omega /k_{B}\sigma _{Q})\sigma _{2}(\omega ,T)$ and universal line $(8/\pi
)T$ [5]( here $\sigma _{Q}=e^{2}/\hbar d$ is quantum conductivity of a stack
of planar conductors with interlayer spacing $d$, $\sigma _{2}(\omega ,T)$
is imaginary part of conductivity $\sigma (\omega ,T)$. It is seen from
fig.2 [3], that this crossover occurs at temperature 62 K, which is limiting
the region of frequency independent conductivity $\sigma _{2}$. As we see
from results of refs.[1,2], the samples with $T_{c}\simeq 74K$ have $\delta
=0.2135$ and $T_{BKT}\simeq 61.72K$. Thus, we can say that at $T\ge 62K$
first unbound vortices appear that leads to the observation of frequency
dependent conductivity $\sigma _{2}(\omega ,T)$ and to exponential
dependence $\rho _{c}(T)$ at $T_{c}<T<95K$ [2].

>From the comparison of the results of refs.[1,2] and [3] it follows that the
region of the temperatures, $62 K <T< 95 K$, where the contribution in
short-timescale dynamics of thermally generated vortices is distinguishable
from another contributions, coincides for both measurements. However in
underdoped HTS, forceful evidences for some form of pairing are found up to
the temperature $T^{*}$ as in spectroscopic measurements [6], so as both in
resistivity measurements [1,2] and in measurements of Knight shift [7-8].
Now the question is under the discussion: which is just form of pairing
without phase coherence in the region $T^{*}>T> 95K$? It can be or d-wave
pairing of normal electrons with momentum for which superconducting gap
vanishes [9], and or pairing in striped $Cu O_2$ planes, which leads to
two-dimensional superconducting fluctuations in the theory Ginzburg-Landau
(see expression (3) and ref.[2]).

References

1. T.Watanabe, T.Fujii, and A.Matsuda, Phys.Rev.Lett. 79, 2113 (1997)

2. G.G.Sergeeva, Low Temp. Phys. 26, 331 (2000)

3. J.Corson, R.Mallozzi, J.Orenstein, J.N.Eckstein and I.Bozovic, Nature,
398, 221 (1999)

4. G.G.Sergeeva, V.Yu.Gonchar, A.V.Voitsenya, Low Temp. Phys.( to be
published)

5.J.M.Kostelitz, and D.R.Nelson, Phys.Rev.Lett. 39, 1201 (1977)

6. Ch.Renner, B.Revaz, J.-Y.Genoud, K.Kadovaki, O.Fisher, Phys.Rev.Lett. 80,
149 (1998)

7. Guo-qing Zheng, et al. Phys.Rev.Lett. 85, 405 (2000)

8. Guo-qing Zheng, et al. Physica B 281-282, 901 (2000)

9. P.A.Lee and X.G.Wen, Phys.Rev.Lett. 78, 4111 (1997)

\end{document}